\documentclass[twocolumn,pra,aps]{revtex4}
\usepackage{graphicx}

\begin{document}

\title{The Free Will Theorem and the Flash Ontology \\ implicitly assume the Before-Before Experiment and thereby imply free will.\\Comment on a note by Nicolas Gisin in arXiv:1002.1392v1}

\author{Antoine Suarez}
\address{Center for Quantum Philosophy, P.O. Box 304, CH-8044 Zurich, Switzerland\\
suarez@leman.ch, www.quantumphil.org}

\date{February 13, 2010}

\begin{abstract} It is argued that both the ``Free Will Theorem'' (FWT) and the ``relativistic GRW model with flash ontology'' (rGRWf) \emph{hiddenly} assume the before-before experiment's result, and for this reason both FWT and rGRWf imply free will in the world outside free experimenters.\\

\end{abstract}

\pacs{03.65.Ta, 03.65.Ud, 03.30.+p, 04.00.00, 03.67.-a}

\maketitle

The Free Will Theorem of John Conway and Simon Kochen \cite{fwt} assumes that experimenters are capable of freely choosing measurement settings.  The theorem basically states that this assumption of free will ``in the particles inside ourselves`` and the experimental violation of Bell's inequality together imply free will also in ``the particles all over the universe''.

In more technical terms, what the authors of the FWT mainly claimed in the first version of their theorem is:
\begin{small}
\begin{eqnarray}\label{1}
  Freedom + Bell+ determinism \Rightarrow contradiction
\end{eqnarray}
\end{small}
Where:

\emph{Freedom} means the supposed capacity of  experimentalist A of choosing his measurement settings independently of the choices done by his colleague experimentalist B.

\emph{Bell} means experiments demonstrating violation of Bell inequalities \cite{jb}.

\emph{Determinism} means that the outcome of an experiment is determined by the information accessible to the particles from the past (\emph{temporal causality}).

Statement (\ref{1}) was first questioned by Roderich Tumulka \cite{tu}. Without entering into the (undoubtedly important) subtleties of his formulations, what Tumulka states is that \emph{Bell experiments} do not imply (\ref{1}) but rather:
\begin{small}
\begin{eqnarray}\label{2}
  Freedom + Bell + locality \Rightarrow contradiction
\end{eqnarray}
\end{small}
In other words, the failure of determinism does not necessarily follow from Bell experiments.

In the modified stronger version of their theorem Conway and Kochen state \cite{sfwt}:

\begin{small}
\begin{eqnarray}\label{3}
  Freedom + Bell + covariant \;deterministic\;dynamics \nonumber\\ \Rightarrow contradiction
\end{eqnarray}
\end{small}
and:
\begin{small}
\begin{eqnarray}\label{4}
  Freedom + Bell+ covariant \;stochastic\;dynamics \nonumber\\ \Rightarrow contradiction
\end{eqnarray}
\end{small}
Where:

\emph{Covariant dynamics} refers to events that are related to each other through only covariant or Lorentzinvariant (relativistic) links. If these events are locally deterministic one has a covariant deterministic model, and if they are locally random a covariant stochastic one.

Regarding statement (\ref{3}) Tumulka and co-authors, and also other authors, claim that it is not new \cite{gold,ng}, but is nothing other that Bell's theorem otherwise formulated. (It seems to me that \cite{ng1} actually provides a simple and elegant proof of (\ref{3})).

By contrast, Tumulka and co-authors consider statement (\ref{4}) wrong. To prove this they give an explicit model called ``relativistic GRW model with flash ontology'', (rGRWf) which according to the authors is stochastic and covariant \cite{gold}.

In a recent comment Nicolas Gisin nicely illustrates
that the origin of this somewhat confused dispute lies in very different understanding of what a ``covariant quantum process'' is. Gisin stresses that in rGRWf the events that happen (each single pair of flashes $f_A$, $f_B$) cannot be covariant. Covariant is only the ``cloud of future events'' (i.e. the probability distribution). Therefore, in some sense, both \cite{sfwt} and \cite{gold} are correct: while Tumulka and co-authors ``correctly insist that rGRWf is as covariant as possible'', Conway and Kochen ``correctly stress that it is not more covariant than possible''\cite{ng}.

The aim of my Comment is to show, \emph{firstly}, that if the content of FWT reduces to statement (\ref{3}), then the theorem does not deserve the name of ``free will'', for it does not exclude non-local determinism. \emph{Additionally}, I show regarding (\ref{4}) that by declaring Flash Ontology (rGRWf) covariant Tumulka and co-authors implicitly assume the result of the before-before (Suarez-Scarani) experiment \cite{asvs97, as00.1, szsg}.

Indeed statement (\ref{3}) implies only the failure of \emph{local} determinism. It is however possible to construct a Suarez-Scarani model that is both non-local and deterministic, i.e., each event is determined by the information accessible from the past although in a non-covariant way \cite{as08}. We remind that the Suarez-Scarani model considers relativistic experiments with beam-splitters in motion in such a way that each of them, in its own reference frame, is first to select the output of the photons (before-before timing). Then, each outcome becomes independent of the other, and the \emph{nonlocal} correlations should disappear \cite{asvs97, as00.1}. Such a non-local model assumes time-ordered non-covariant influences, and if one combines it with local determinism one obtains a nonlocal full-deterministic (temporal causal) model. For before-before timing the model predicts disappearance of nonlocal correlations with maintenance of possible local ones \cite{as09}, and does not entail signaling \cite{vang}. Suarez-Scarani models have been experimentally tested and refuted \cite{szsg}.

Consequently, a more appropriate version of the \emph{Free Will} theorem would be:

\begin{small}
\begin{eqnarray}\label{5}
  Freedom + Suarez\& Scarani
  &+& determinism \nonumber\\&\Rightarrow& contradiction
\end{eqnarray}
\end{small}
In any case, (\ref{5}) excludes more determinism, and thereby allows for more free will than (\ref{1}) and (\ref{3}).

We turn now to statement (\ref{4}): How is it possible that Tumulka and co-authors claim that the Flash Ontology is covariant \cite{gold} (and not only ``as covariant as possible'' like Gisin suggests \cite{ng})? It seems to me that the reason for this claim is hidden in the following Tumulka's statement in \cite{tu}: ``The objective facts are where-when the flashes occur, and it is enough if a theory prescribes, as does rGRWf, their joint distribution in a Lorentzinvariant way. Whether nature chooses the space-time point $f_B$ first, and $f_A$ afterwards, or the other way around, does not seem like a meaningful question to me.''

As said above, it is possible to construct a testable, and hence physically ``meaningfull'' model assuming time-ordered links between space-like separated events. Thus the question Tumulka refers to is clearly ``meaningful'', and he himself is wrong. Yet the before-before experiment proves Suarez-Scarani models wrong and demonstrates that nature does the correlations dismissing any time order. In this sense Tumulka is right: On the one hand, if the time order does not exist, then each single pair of flashes ($f_B$,$f_A$) can be considered a single event coming from outside space-time, and the concept of covariance does not apply. On the other hand, there where this concept makes sense, the model is covariant. In conclusion, it is before-before experiment which makes Flash Ontology covariant.

But there is something more. Demonstrating local randomness alone, though it suffices to refute determinism, it is not yet enough to prove ``free will''. Indeed anyone who believes to share ``this valuable commodity'' will certainly expect to be able of controlling surrounding randomness to some extent. Now the before-before experiment demonstrates effects in which control of quantum randomness happens from outside space-time, and in this sense it can also be considered an experimental proof of free will on the part of nature (provided there is free will in our brains!) \cite{as08.1}. But this also means that Flash Ontology implies free will since it assumes \emph{correlated} ''flashes`` that cannot be explained by any history in space-time.

In conclusion, no experiment can prove ``the free will of the experimenter''. But if one assumes this free will, then the before-before experiment demonstrates free will also in the world outside the experimenter. Since the FWT and the rGRWf Flash Ontology implicitly assume the before-before experiment, both descriptions imply free will in ``the particles all over the universe'' and, in this respect, FWT and the rGRWf are equivalent.

\noindent\emph{Acknowledgments}: I am grateful to Nicolas Gisin  for comments on the draft.

\end{document}